\documentclass[pdflatex,sn-mathphys]{sn-jnl}

\usepackage{amsmath}
\usepackage{amssymb}
\usepackage{graphicx}
\usepackage{subcaption}
\usepackage{url}
\usepackage{color}

\newtheorem{lemma}  {Lemma} 
\newtheorem{theorem}  {Theorem} 
\renewcommand{\qed}{\hfill $\Box$}

\begin{document}


\title[Navigating Planar Topologies in Near-Optimal Space and Time]{Navigating Planar Topologies in Near-Optimal Space and Time\footnote{Funded by ANID - Millennium Science Initiative Program - Code ICN17\_002, Chile and by European Union's Horizon 2020 research and innovation programme under the Marie Sklodowska-Curie grant agreement No 690941 (project BIRDS). The authors received funding from Fondecyt grants 77190038, 1200038, and 1170497, respectively. An early partial version of this paper appeared in {\em Proc. SPIRE 2019}.}}





\author[1]{\fnm{Jos\'e} \sur{ Fuentes-Sep\'ulveda}}\email{jfuentess@udec.cl}
\equalcont{These authors contributed equally to this work.}

\author*[2,3]{\fnm{Gonzalo} \sur{Navarro}}\email{gnavarro@dcc.uchile.cl}

\author[1,3]{\fnm{Diego} \sur{Seco}}\email{dseco@udec.cl}
\equalcont{These authors contributed equally to this work.}

\affil[1]{\orgdiv{Dept. of Computer Science}, \orgname{Universidad de Concepci\'on}, \orgaddress{\city{Concepci\'on}, \country{Chile}}}

\affil[2]{\orgdiv{Dept. of Computer Science}, \orgname{University of Chile}, \orgaddress{\city{Santiago}, \country{Chile}}}

\affil[3]{\orgname{Millennium Institute for Foundational Research on Data (IMFD)}, \orgaddress{\city{Santiago}, \country{Chile}}}

\abstract{
We show that any embedding of a planar graph can be encoded succinctly while efficiently answering a number of 
topological queries near-optimally. More precisely, we build on a succinct 
representation that encodes an embedding of $m$ edges within
$4m$ bits, which is close to the information-theoretic lower bound of about $3.58m$. 
With $4m+o(m)$ bits of space, we show how to answer a number of topological queries relating
nodes, edges, and faces, most of them in any time in $\omega(1)$. Further, we 
show that with $O(m)$ bits of space we can solve all those operations in 
$O(1)$ time.}


\maketitle

\section{Introduction}
Plane embeddings, which are drawings of planar graphs on the plane, arise naturally in many applications, especially in those that are geometrical in nature like VLSI, computer graphics, and Geographic Information Systems (GIS) \cite{LozzoDF20}. In this work we focus on efficiently answering queries that relate nodes, edges, and faces in planar embeddings. Those are the building blocks, for example, of the topological model, widely used in GIS applications to describe topological relationships among objects. With this underlying motivation, we define a comprehensive set of topological queries and show that they can be efficiently answered within very little space.

To achieve such space-efficient representations, we build on compact data structures (CDS) \cite{Navarro2016}, whose main goal is to support efficient query operations while using space close to the information-theoretic lower bound. CDS have achieved remarkable results, both in theory and practice, to handle very large volumes of data in different domains, including graph and geometric data~\cite{aleardi2005succinct,aleardi2008succinct,BoseCHMM12}.

We build on Tur\'an's encoding \cite{Turan1984} of plane embeddings of connected 
planar graphs, where the {\em dual} of the graph (where the faces become nodes)
is also explicitly represented. We build on that representation and extend
previous results \cite{FFSGHN18} in order to provide a succinct-space 
representation of the topological model (using $4m+o(m)$ bits on a graph
of $m$ edges) that efficiently
supports a rich set of topological queries (most of them in any time in $\omega(1)$), which include those defined in current standards and flagship
implementations. Our main technical results are new 
$\omega(1)$-time algorithms for determining if two nodes
are neighbors, and if a node touches a face, by orienting edges; many other results are derived via analogous structures and
exploiting duality. 

We then improve the time complexities by relaxing the space usage to $O(m)$
bits. We show that, within this space, all the operations 
can be supported in $O(1)$ time. Our main technique in this 
second part is a new 
$O(m)$-bits representation that allows us determining in constant time whether two nodes are in touch with the same unknown face.

\section{Our Contribution in Context}

\subsection{An Application in GIS: The Topological Model}
\label{subsec:topomodel}

Geographic Information Systems (GIS) enable \emph{capture, modeling, manipulation, retrieval, analysis and presentation}~\cite{Worboys:2004} of geographically referenced data. On the logical level, the most popular GIS model (together with the raster model) is the vector model. There are three common representations of collections of vector objects, called \emph{spaghetti}, \emph{network}, and \emph{topological} model, which mainly differ in the expression of topological relationships among the objects~\cite{2001:SDA}. In the spaghetti model, the geometry of each object is represented independently of the others and no explicit topological relations are stored. Despite its drawbacks, this is the most used model in practice because of its simplicity and the lack of efficient implementations of the other models. Those other two models are similar, and explicitly store topological relationships among objects. The network model is tailored to graph-based applications, such as transportation networks, whereas the 
topological model focuses on planar networks (e.g., all sorts of maps).
This model is more efficient to answer topological queries, which are 
usually expensive, thus it is gaining popularity in spatial databases like Oracle Spatial.

The topological model represents a planar subdivision into adjacent
polygons. Hereinafter, we will refer to these polygons as \emph{faces}. A face
is represented as a sequence of \emph{edges}, each of them being shared with an
adjacent face, which may be the outer face. An edge connects two 
\emph{nodes}, which are associated with a point in space, usually the
Euclidean space. Edges also have a geometry, which represents the boundary
shared between its two faces. This eliminates redundancy in the stored 
geometries and also reduces inconsistencies. 
In Fig.~\ref{fig:succrep},
faces are named with capital letters, $A$ to $I$, $A$ being the outer face. Face
$C$ is defined by the sequence of nodes $\langle 1,5,4,8,7,6 \rangle$, and edge 
$(6,7)$ is shared by faces $C$ and $F$.
Note, however, that a pair of nodes is insufficient in general to name an edge,
because multiple edges may exist between two nodes.

Those topological concepts are related with geographic entities. The basic
geographic entity is the point, defined by two coordinates. Each node 
in the topological model is associated with a point, and each edge is
associated with a sequence of points describing a sequence of segments that
form the boundary between the two faces that share such edge. Each face is
related to the area limited by its edges (the external face is infinite).

The international standard ISO/IEC 13249-3:2016~\cite{isosqlmm} defines a basic
set of primitive operations for the model, which are also implemented in
flagship database systems\footnote{\url{http://postgis.net/docs/Topology.html}}.
Some of the queries relate the geometry with the topology, for example, find 
the face covering a point given its coordinates. Those queries require data 
structures that store coordinates, and are therefore bound to use considerable
space. Instead, we focus on {\em pure topological} queries, which can be solved
within much less space and can encompass many problems once mapped to topological 
space. We also restrict our work to a static version of the model, in which 
case our representation supports a much richer set of access operations.  

Topological queries can also be solved using the geometries, but this approach 
is computationally very expensive. We propose instead an approach in which
most of the work is done on an in-memory compact index on the topology,
resorting to the geometric data only when necessary. Such an approach
enables handling geometries that do not fit in main memory, but whose
topologies do, and still solving queries on them with reasonable efficiency
because secondary-memory accesses are limited.
To illustrate this, consider the example of \emph{given the coordinates of 
two query points, tell
if they lie on adjacent faces, and if so, which edge separates them}. In our
approach, this type of query can be solved with just two mappings from the
geographical space to the topological space, and then using pure topological 
queries.

Table~\ref{tab:model} lists a set of topological queries we consider on the 
topological model, together with the time complexities we achieve in this 
paper within $4m+o(m)$ bits (we solve them all in $O(1)$ time within $O(m)$
bits). These comprehensively consider querying about
relations between two given entities of the same or different type, and listing
or counting entities related to a given one. The set considerably extends the
queries available in standards or flagship implementations, which
comprise just {\tt intersects} (1.d and 5.b), {\tt GetNodeEdges} (3.c), and
{\tt ST\_GetFaceEdges} (3.e). 

\begin{table}[t]
\caption{The queries we consider on the topological model and our time complexities
within $4m+o(m)$-bit space. We put in boldface those where we contributed.}
\label{tab:model}
\begin{center}
\vspace{5mm}
\begin{tabular}{|@{~}c@{~}l@{~}|@{~}c@{~}|@{~}l@{~}|}
\hline
\multicolumn{4}{|c|}{1. Relations between entities of the same type} \\
\hline
{\bf (1.a)} & Do edges $e$ and $e'$ share a node? 
	& $O(1)$
	& \cite{FFSGHN18} $+$ Lemma~\ref{lem:nodeface} \\
{\bf (1.b)} & Do edges $e$ and $e'$ border the same face?
	& $O(1)$
	& \cite{FFSGHN18} $+$ Lemma~\ref{lem:nodeface} \\
{\bf (1.c)} & Do nodes $u$ and $v$ share an edge?
	& any in $\omega(1)$
	& Lemma~\ref{lem:neigh} \\
{\bf (1.d)} & Do faces $x$ and $y$ share an edge? 
	& any in $\omega(1)$
	& Lemma~\ref{lem:dualneigh} \\
\hline
\multicolumn{4}{|c|}{2. Relations between entities of different type} \\
\hline
{\bf (2.a)} & Is edge $e$ incident on node $u$? 
	& $O(1)$
	& \cite{FFSGHN18} $+$ Lemma~\ref{lem:nodeface} \\
{\bf (2.b)} & Is edge $e$ on the border of face $x$?
	& $O(1)$
	& \cite{FFSGHN18} $+$ Lemma~\ref{lem:nodeface} \\
{\bf (2.c)} & Is face $x$ incident on node $u$?	
	& any in $\omega(1)$ 
	& Lemma~\ref{lem:nodesfaces} \\
\hline
\multicolumn{4}{|c|}{3. Listing related entities (time per element output)} \\
\hline
{\bf (3.a)} & Endpoints of edge $e$
	& $O(1)$ 
	& \cite{FFSGHN18} $+$ Lemma~\ref{lem:nodeface} \\
{\bf (3.b)} & Faces divided by edge $e$
	& $O(1)$ 
	& \cite{FFSGHN18} $+$ Lemma~\ref{lem:nodeface} \\
(3.c) & Nodes/edges neighbors of node $u$
	& $O(1)$ 
	& \cite{FFSGHN18} \\
(3.d) & Faces bordering face $x$
	& $O(1)$ 
	& \cite{FFSGHN18} and duality \\
{\bf (3.e)} & Faces incident on node $u$
	& $O(1)$ 
	& Lemma~\ref{lem:listnodesorfaces} \\
{\bf (3.f)} & Nodes/edges bordering face $x$
	& $O(1)$ 
	& Lemma~\ref{lem:listnodesorfaces} \\
\hline
\multicolumn{4}{|c|}{4. Counting related entities} \\ 
\hline
{\bf (4.a)} & Nodes/edges/faces neighbors of node $u$
	& any in $\omega(1)$
	& \cite{FFSGHN18} extended \\
{\bf (4.b)} & Faces/edges/nodes bordering face $x$
	& any in $\omega(1)$ 
	& \cite{FFSGHN18} and duality \\
\hline
\multicolumn{4}{|c|}{5. Relations via a third entity} \\
\hline
{\bf (5.a)} & Do nodes $u$ and $v$ border the same face?
	& any in $\omega(\sqrt{m})$
	& Lemma~\ref{lem:sqrt} \\
{\bf (5.b)} & Do faces $x$ and $y$ share a node? 
	& any in $\omega(\sqrt{m})$
	& Lemma~\ref{lem:sqrt} \\
\hline
\end{tabular}
\end{center}
\end{table}

\subsection{Planar Graphs}

A graph is {\em planar} if it can be drawn on the plane without crossing
edges. The topology of a specific drawing of a planar graph on the plane is 
called a {\em plane embedding}. We use plane embeddings to represent 
topological models. Representing a plane embedding with $m$ edges requires 
$m\log 12 \approx 3.58m$ bits \cite{Tutte1963}, which opens the door to
$O(m)$-bit representations. This is remarkable because representing general 
graphs with
$n$ nodes and $m$ edges needs $\Omega(m\log\frac{n^2}{m})$ bits.

Succinct representations of plane embeddings build on spanning trees, book 
embeddings~\cite{Yannakakis1979}, and small node separators~\cite{lt1979}.
Tur\'an~\cite{Turan1984} introduced a succinct representation using $4m$ bits, 
and Keeler and Westbrook~\cite{KeelerWestbrook1995} reached the optimal 
$m\log 12 + O(1)$ bits, though disallowing either self-loops or nodes with 
degree 1. Both used spanning trees. He et al.~\cite{HKL00} used graph separators
and obtained $m\log 12 + o(m)$ bits without restrictions. Those representations
do not support efficient navigation of the compressed representation, however.

There exist a number of navigable representations, which support a few basic queries in optimal time: adjacency (are these two nodes connected?), degree 
(how many neighbors this node has?), and neighborhood (list the neighbors of 
this node, in clockwise or counter-clockwise order).
Using book embeddings, Jacobson~\cite{Jacobson1989} provides an $O(m)$-bit 
representation, and Munro and Raman~\cite{MR01} provide a $2m+8n+o(m)$ bits 
representation not allowing self-loops. Blelloch and 
Farzan~\cite{BlellochFarzan2010} propose a representation using $m\log 12+o(m)$
bits based on small node separators. 
Ferres et al.~\cite{FFSGHN18} use spanning trees to provide a simple 
representation using $4m+o(m)$ bits with richer functionality.

There are other succinct representations of planar graphs, but most of them 
cannot represent an arbitrary embedding either because they need to change the 
embedding in order 
to compute spanning trees with useful properties, or because they add 
edges to the embedding in order to obtain a planar triangulation. We refer the 
reader to Navarro's book~\cite[Sec.~9.4]{Navarro2016} for more details.

\subsection{Our Contribution}

In this paper we are interested in the representation of Ferres et 
al.~\cite{FFSGHN18}, which extends Tur\'an's encoding with $o(m)$ extra 
bits in order to support efficient navigation operations. They list neighbors
in optimal time, and show how to list all the edges of a face in optimal time
as well. However, computing degrees requires (any) time in $\omega(1)$ and
determining adjacency of two nodes requires (any) time in $\omega(\log n)$. 
Compared to other more efficient representations, however, Tur\'an's encoding
is interesting because it includes an explicit representation of both the 
plane embedding and of its dual, that is, one can directly refer to faces and pose queries on them. We use this feature
to extend the set of primitives so as to support a full set of topological queries, formed by 
all the operations listed in Table~\ref{tab:model}. Moreover, we
improve their performance for adjacency and related queries to any time in $\omega(1)$.

A warmup result essentially hinted by Ferres et al., our
Lemma~\ref{lem:nodeface}, sorts out a number of simple queries (all [123].[ab])
in constant time. A consequence of Lemma~\ref{lem:nodeface} is 
Lemma~\ref{lem:listnodesorfaces}, which
extends the algorithm of Ferres et al.\ listing the neighbors of a node 
(3.c, {\tt GetNodeEdges}) in optimal time to list the faces incident on a node
(3.e) and, by duality, to list the faces or edges bordering a face (3.d,
{\tt ST\_GetFaceEdges}) and the nodes bordering a face (3.f), all in optimal
time. We also extend their results that count the edges 
incident on a node (4.a) in $\omega(1)$ time to count nodes, edges,
or faces incident on a node or bordering a face (4.b).

Our first main result is Lemma~\ref{lem:neigh}, which exploits orientation of edges 
to determine if two given nodes are connected by an edge (1.c) in any time in $\omega(1)$, adding only $o(m)$ bits to the main
structure. The same procedure on the dual graph, Lemma~\ref{lem:dualneigh},
determines in the same time if two given faces share an edge (1.d, a variant 
of the standard query \texttt{intersects}). Our second main result is Lemma~\ref{lem:nodesfaces}, which builds on Lemma~\ref{lem:neigh} to determine if a given node is in the frontier of a 
given face (2.c) in any time in $\omega(1)$, by defining a new graph where faces become nodes as well. Determining if two 
given nodes border the same face (5.a) or if two given faces share some node 
(5.b, a variant of query {\tt intersects}) is costlier, 
$\omega(\sqrt{m})$. 

Our general approach is to solve the queries by enumeration (queries of type 3), mapping the ``hard'' nodes/faces where enumeration would be too expensive to a smaller graph where we can store extra information in $o(m)$ bits that allows handling the query in $O(1)$ additional time. The main challenge is to define what extra information to store, and how to store it, so that we take $o(m)$ bits of space and we can map to the reduced graph in constant time.

If we use $(4+\epsilon)m$ bits of space, for any constant
$\epsilon > 0$, then we have space to interpret ``too expensive'' in the previous paragraph to ``more than a constant'', then all the times of the form `` any in $\omega(1)$'' in Table~\ref{tab:model} become
$O(1)$. We then explore what can be achieved if we allow any space usage in
$O(m)$ bits.  Lemma~\ref{lem:inters-constant} 
shows that, by modifying the arrangement of Lemma~\ref{lem:nodesfaces},
we also solve queries (5.a) and (5.b) in constant time. 
As a result, all the queries in Table~\ref{tab:model} can be solved in $O(1)$ 
time and within $O(m)$ bits of space.

The following theorem summarizes our results.

\begin{theorem}
An embedding of a connected planar graph with $m$ edges can be represented in $4m+o(m)$ bits so that the queries listed in Table~\ref{tab:model} can be answered in the given time complexities. By using $O(m)$ bits, all the queries can be solved in $O(1)$ time. 
\end{theorem}

Note that the given space results assume that the graph is connected. Ferres et al. 
\cite{FFSGHN18} show how an embedding formed by $k$ connected components can be 
optimally represented by adding $k\lg(m/k)+O(k)$ bits and without any essential
change to the algorithms designed for connected graphs.

A preliminary version of this article appeared in {\em Proc. SPIRE'19}
\cite{FNS19}. In this extended version we present the results in greater 
detail, and manage to improve their time for queries (1.c) and (1.d) from $O(\frac{\log\log m}{\log\log\log m})$, and for query (2.c) from $\omega(\log n)$, to any time in $\omega(1)$. Further, we obtain constant time on all the queries of
Table~\ref{tab:model} by relaxing the space usage to $O(m)$ bits.

\section{Succinct Data Structures}

\subsection{Sequences and Parentheses}

Given a sequence $S[1..n]$, the
operation $rank_{a}(S,i)$ returns the number of occurrences of the symbol
$a$ in the prefix $S[1..i]$, and the operation $select_{a}(S,i)$ returns
the position in $S$ of the $i$th occurrence of the symbol $a$. For binary
alphabets, the bitvector $S$ can be stored in $n+o(n)$ bits supporting $rank$ and 
$select$ in $O(1)$ time~\cite{Cla96}. If $S$ has $m$ 1-bits, then it can be
represented in $m\lg\frac{n}{m}+O(m)+o(n)$ bits, maintaining $O(1)$-time
$rank$ and $select$~\cite{RRR07}. 

Binary sequences can be used to represent balanced parentheses
sequences, by interpreting the bit values as opening or closing parentheses.
Given a balanced parenthesis sequence $S$, $open(S,i)/close(S,i)$
returns the position in $S$ of the closing/opening parenthesis matching the
parenthesis $S[i]$, and $enclose(S,i)$ returns the rightmost position $j$ such
that $j< i< close(S,j)$. A parentheses sequence can be used to represent 
an ordinal tree, where each node is identified by the position $i$ of an opening
parenthesis $S[i]$ and its descendant nodes are listed between positions $i+1$
and $close(S,i)-1$. The parent of the node $i$ is
$parent(S,i)=enclose(S,i)$. Another relevant operation for this interpretation
is $child(S,i,j)$, which yields the opening parenthesis of the $j$th child of
the node identified by position $i$. The sequence $S$ can be represented in 
$n+o(n)$ bits and support $open$, $close$, $enclose$, $parent$, and $child$,
all in $O(1)$ time~\cite{NS14}. 

\subsection{Ferres et al.'s Representation}
\label{sec:ferres}

\begin{figure}[t]
    \centering
    \begin{subfigure}[b]{0.45\textwidth}
        \includegraphics[width=.75\textwidth]{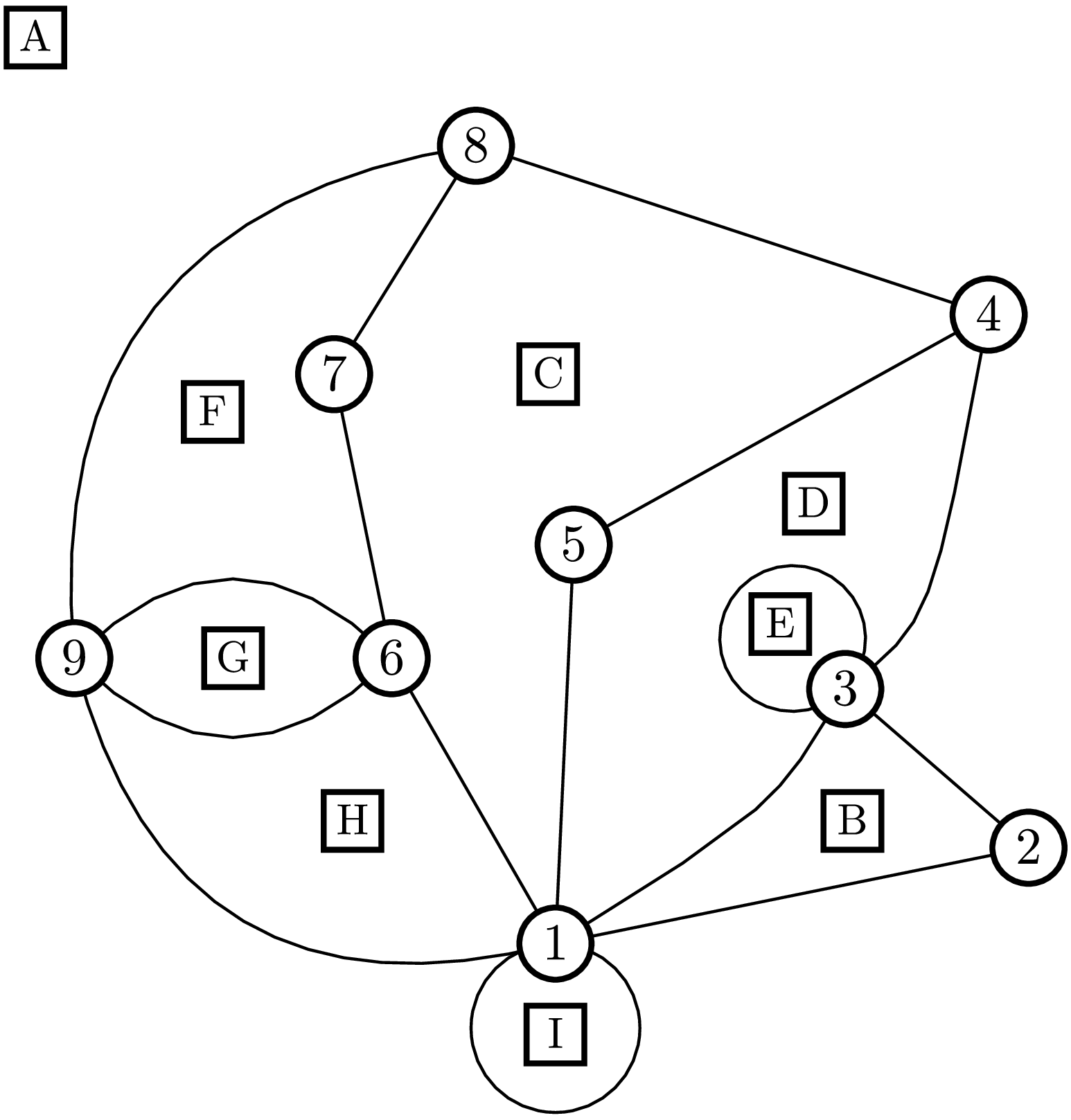}
        \caption{A plane embedding}
        \label{fig:planar}
    \end{subfigure}
    ~ 
    \begin{subfigure}[b]{0.45\textwidth}
        \includegraphics[width=.95\textwidth]{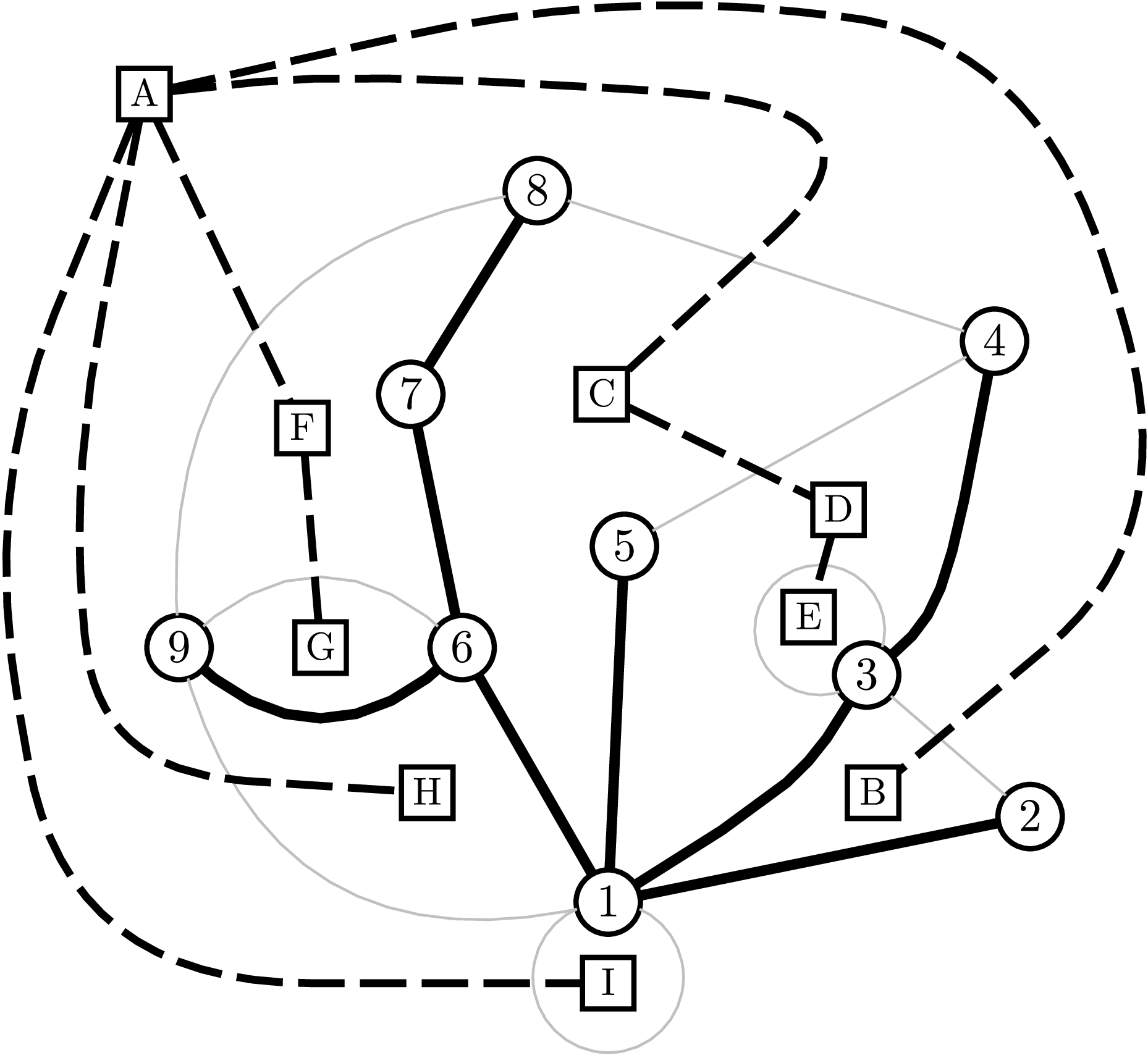}
        \caption{Its primal and dual spanning trees}
        \label{fig:primal-dual}
    \end{subfigure}
    \begin{subfigure}[b]{0.9\textwidth}
        \includegraphics[width=\textwidth]{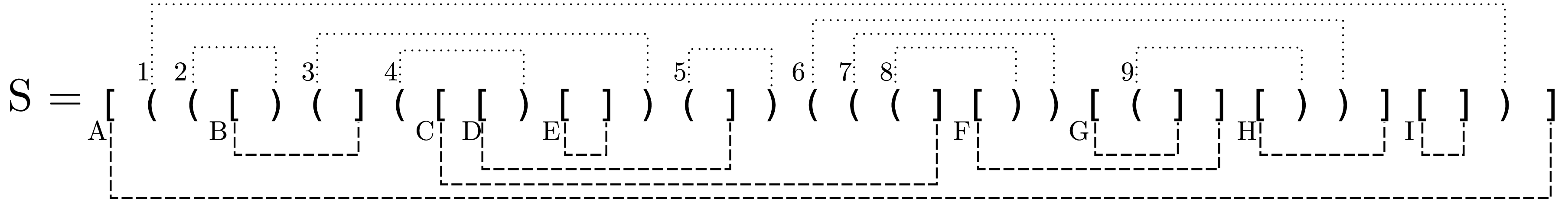}
        \caption{The sequence of parentheses and brackets encoding the plane embedding}
        \label{fig:sequence}
    \end{subfigure}
    \begin{subfigure}[b]{0.9\textwidth}
        \includegraphics[width=\textwidth]{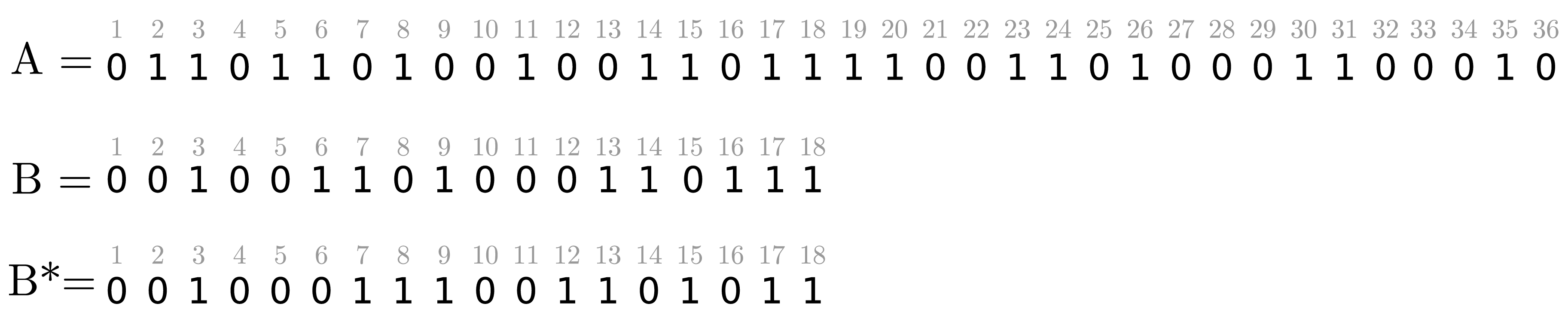}
        \caption{Bitvectors $A$, $B$ and $B^*$  representing the sequence $S$}
        \label{fig:sequencesAB}
    \end{subfigure}
    \caption{Example of the succinct plane embedding representation of Ferres
      et al.~\cite{FFSGHN18}.}\label{fig:succrep}
\end{figure}

Given a plane embedding of a connected planar graph $G$, the computation of a spanning
tree $T$ of $G$ induces a spanning tree $T^*$ in the dual graph of
$G$~\cite{Biggs1971}. 
The edges of $T^*$ correspond to the edges in the dual graph that cross 
edges in $G\setminus T$. 
Fig.~\ref{fig:primal-dual} shows a primal (thick continuous edges) and a dual
(thick dashed edges) spanning trees for the plane embedding of 
Fig.~\ref{fig:planar}. Lemma~\ref{lem:travis} states that
a depth-first traversal of $T$ induces a depth-first traversal in $T^*$.

\begin{lemma}[{\cite{FFSGHN18}}] \label{lem:travis}
Consider any plane embedding of a planar graph $G$, any spanning tree $T$ of  
$G$ and the complementary spanning tree $T^*$ of the dual $G^*$ of $G$. 
Suppose we 
perform a depth-first traversal of $T$ starting from any node on the outer
face of $G$ and always process the edges incident to the node v we are 
visiting in counter-clockwise order. At the root, we arbitrarily choose an 
incidence of the outer face in the root and start from the last edge of the 
incidence in counterclockwise order; at any other node, we start from the 
edge immediately after the one to that node's parent. Then each edge not in 
$T$ corresponds to the next edge we cross in a depth-first traversal of $T^*$.
\end{lemma}

Here, an incidence of the outer face in the root means a place
where the root and the outer face are in contact.
For instance, in Fig.~\ref{fig:primal-dual}, the traversal
can start at edge $(1,1)$, $(1,2)$, or $(1,9)$, taking node
$1$ as the spanning tree root.

The compact representation \cite{Turan1984,FFSGHN18} is based on the traversal
of Lemma~\ref{lem:travis}. Starting at the root of any suitable spanning tree 
$T$, each time we visit for the first time an edge $e$, we write a ``('' if $e$ 
belongs to $T$, or a ``['' if not. Each time we visit an edge $e$ for the 
second time, we write a ``)'' if $e$ belongs to $T$ or a ``]'' otherwise.
We call $S$ the resulting sequence of $2m$ parentheses and brackets, which are
enclosed by an additional pair of parentheses and of brackets that represent
the root and the outer face, respectively.
Ranks of opening parentheses act as node identifiers,
whereas ranks of opening brackets act as face identifiers. 
Further, positions in $S$ act as edge identifiers: each edge is identified twice, first by
an opening parenthesis or bracket, and later by its corresponding closing
parenthesis or bracket.

Fig.~\ref{fig:sequence} shows the sequence $S$ for the plane embedding of
Fig.~\ref{fig:primal-dual}, starting the traversal at the edge
$(1,2)$. Observe that the parentheses of $S$ encode the balanced-parentheses
representation of $T$ and the brackets encode the balanced-parentheses 
representation of the dual spanning tree $T^*$. In general the representation
of a node $v$, $( \cdots )$ where $``(''$ is the $v$th opening parenthesis, 
contains the sequences of nested parenthesis sequences for the children of
$v$ in $T$ (e.g., the sequence for node $1$ in the figure contains those
of the nodes $2$ and $8$), interspersed with top-level brackets (i.e., brackets not
contained in the sequence of a child of $v$). Those brackets represent the
other edges incident on $v$ (e.g., the $``]''$ of G and the $``[~]''$ of H).
Since the brackets also represent the faces of $G$,
if we exchange the roles of brackets and parentheses, the sequence represents
the dual graph $G^*$.

In the succinct representation of Ferres {\it 
et al.}~\cite{FFSGHN18}, the sequence $S$ is stored in three bitvectors,
$A[1..2(m+2)]$, $B[1..2n]$, and $B^*[1..2(m-n+2)]$. It holds that $A[i]=1$ if 
the $i$th entry of $S$ is a parenthesis, and $A[i]=0$ if it is a bracket. 
Bitvector $B$ stores the balanced sequence of parentheses of $S$, storing a $0$
for each opening parenthesis and a $1$ for each closing parenthesis. Bitvector
$B^*$ stores the balanced sequence of brackets of $S$ in a similar way. Fig.~\ref{fig:sequencesAB} shows the bitvectors that store the sequence $S$ of Fig.~\ref{fig:sequence}.

Adding support for $rank$ and $select$ operations on $A$, $B$ and $B^*$, and for
$open$, $close$ and $enclose$ (i.e., $parent$) operations on $B$ and $B^*$, 
we simulate their support on $S$, as follows ($match(S,i)$ and
$enclose(S,i)$ give the parenthesis or bracket matching or enclosing, 
respectively, the one at $S[i]$):
\[ \small
S[i] = 
	\left\{
	\begin{array}{ll}
	   ``('' & \textrm{if } A[i]=1 \land B[rank_1(A,i)] = 0, \\
	   ``)'' & \textrm{if } A[i]=1 \land B[rank_1(A,i)] = 1 \\
	   ``['' & \textrm{if } A[i]=0 \land B^*[rank_0(A,i)] = 0, \\
	   ``]'' & \textrm{if } A[i]=0 \land B^*[rank_0(A,i)] = 1 \\
	\end{array}
	\right. 
\]
\[ \small
rank_c(S,i)  = 
	\left\{
	\begin{array}{ll}
	   rank_0(B,rank_1(A,i)) & \textrm{if~}c = ``('', \\
	   rank_1(B,rank_1(A,i)) & \textrm{if~}c = ``)'' \\
	   rank_0(B^*,rank_0(A,i)) & \textrm{if~}c = ``['', \\
	   rank_1(B^*,rank_0(A,i)) & \textrm{if~}c = ``]'' \\
	\end{array}
	\right. 
\]
\[ \small
select_c(S,i) = 
	\left\{
	\begin{array}{ll}
	   select_1(A,select_0(B,i)) & \textrm{if~}c = ``('', \\
	   select_1(A,select_1(B,i)) & \textrm{if~}c = ``)'' \\
	   select_0(A,select_0(B^*,i)) & \textrm{if~}c = ``['', \\
	   select_0(A,select_1(B^*,i)) & \textrm{if~}c = ``]'' \\
	\end{array}
	\right. 
\]
\[ \small
match(S,i) = 
	\left\{
	\begin{array}{llll}
	   select_1(close(B,rank_1(A,i))) & \textrm{if~} S[i]=``('', \\
	   select_1(open(B,rank_1(A,i))) & \textrm{if~} S[i]=``)'', \\
	   select_0(close(B^*,rank_0(A,i))) & \textrm{if~} S[i]=``['', \\
	   select_0(open(B^*,rank_0(A,i))) & \textrm{if~} S[i]=``]'' \\
	\end{array}
	\right. 
\]
\[ \small
enclose(S,i) =
	\left\{
	\begin{array}{ll}
	   select_1(enclose(B,rank_1(A,i))) & \textrm{if } S[i]=``('' \\
	   select_0(enclose(B^*,rank_0(A,i))) & \textrm{if } S[i]=``['' \\
	\end{array}
	\right. 
\]

With those primitives, the succinct representation of Ferres et 
al.~\cite{FFSGHN18} supports constant-time operations to navigate the 
embedding. Precisely, the representation supports the following operations:
\begin{description}
\item[$first(v)/last(v)$:] the position in $S$ of the first/last visited edge 
of the node $v$; 
\item[$mate(i)$:] the position in $S$ of the other occurrence of the $i$th
visited edge; 
\item[$next(i)/prev(i)$:] the position of the next/previous edge after
visiting the $i$th edge of a node $v$ in counter-clockwise order; and
\item[$node(i)$:] the index of the node we are visiting when the traversal reaches the $i$th edge. 
\end{description}

The index $v$ of the nodes corresponds to their order in the depth-first 
traversal of the spanning tree $T$, whereas the index $i$ of a visited edge is
just a position in $S$ (since each edge is visited twice, it has two positions
in $S$, and $node(\cdot)$ is different on them). The operations are then supported as follows \cite{FFSGHN18}:

\begin{itemize}
\item
By Lemma~\ref{lem:travis}, the first visited edge of a node $v$ is
immediately after the edge to the parent of $v$ in $T$ (except for the
root of $T$), thus $first(v)=select_{``(''}(S,v) + 1$. The last visited
edge of $v$ is the one returning to its parent, 
$last(v)=match(S,select_{``(''}(S,v))$.
\item
The operation $mate(i)$ is just $match(S,i)$.
\item
The implementation of $next(i)$ depends on whether the $i$th visited edge 
belongs to $T$ or not. Specifically, $next(i)=i+1$ unless $S[i]=``(''$,
in which case it is instead $next(i)=match(S,i)+1$. Analogously,
$prev(i)=i-1$ unless $S[i-1]=``)''$, in which case it is $match(S,i-1)$.
\item
Operation $node(i)$ also depends on whether $S[i]$ is a parenthesis or a 
bracket. In the first case, the edge is in $T$ and connects $i$ with its 
parent. The source is the parent, $rank_{``(''}(S,enclose(S,i))$, if $S[i]=``(''$, or the mate of
$i$, $rank_{``(''}(S,match(S,i))$, if $S[i]=``)''$.
On brackets (i.e., $S[i]= ``[''$ or $``]''$), we must find the lowest node of 
$T$ containing the bracket. That is, letting $j=select_1(A,rank_1(A,i))$
be the position of the last parenthesis preceding $i$,
$node(i)=rank_{``(''}(S,j)$ if $S[j] = ``(''$, and
$node(i)=rank_{``(''}(S,enclose(S,match(S,j)))$ otherwise.
\end{itemize}

With the operations described above, we can implement more complex queries
in optimal time, such as listing all the incident edges of a node $v$ in 
constant time per returned element, and listing all the edges or nodes 
bordering a face given an edge of
the face, spending constant time per returned element. Other operations,
such as the degree of a node and checking if two nodes are
neighbors, are not supported in constant time. For the degree of a node $v$,
the representation supports any time in $\omega(1)$,
whereas for the adjacency test of two nodes $u$ and $v$,
they achieve any time in $\omega(\log m)$. 
Theorem~\ref{thm:basic} summarizes the results of Ferres et al.

\begin{theorem}[\cite{FFSGHN18}] \label{thm:basic}
An embedding of a connected planar graph with $m$ edges can be represented
in $4m + o(m)$ bits, supporting the listing in
clockwise or counter-clockwise order of the neighbors of a node and the nodes
bordering a face in $O(1)$ time per returned node. One can also find the 
degree of a node in any time in $\omega(1)$, and check if two nodes are adjacent in any time in $\omega(\log m)$.
\end{theorem}

\section{Some Simple Results}

As a warm-up exercise, we start with some results that derive easily from
previous work \cite{FFSGHN18}, but that have not been clearly stated.

\subsection{Nodes and Faces Connected by an Edge}

We first obtain the nodes connected by a given edge, and its dual, 
the faces separated by the edge. This trivially answers queries (1.a) and its 
dual (1.b), (2.a) and its dual (2.b), (3.a) and its dual (3.b), all in constant
time.

Note that our edge representation, as positions in $S$, is valid for both
$G$ and its dual $G^*$: by Lemma~\ref{lem:travis}, the spanning tree edges of 
$G$, marked with parentheses in $S$, are exactly the non-spanning tree edges 
of $G^*$, and vice versa, the brackets in $S$ are the spanning-tree edges of 
$G^*$ and the non-spanning tree edges of $G$. 
We then define a new operation, $face(i)$, that returns the identifier of the face where the traversal of $G^*$ is when we traverse edge $i$. This is solved analogously to $node(i)$, by exchanging the meaning of parentheses and brackets:
\begin{itemize}
    \item If $S[i]$ is a bracket, then $face(i)=rank_{``[''}(S,enclose(S,i))$ if $S[i]=``[''$, and $face(i)=rank_{``[''}(S,match(S,i))$ if $S[i]=``]''$.
If $S[i]$ is a parenthesis, we compute the position $j=select_0(A,rank_0(A,i))$ of
the last bracket preceding $i$; then
$face(i)=rank_{``[''}(S,j)$ if $S[j] = ``[''$, and
$face(i)=rank_{``[''}(S,enclose(S,match(S,j)))$ otherwise.
\end{itemize}

The result is trivial once we can compute operations $node(\cdot)$ and $face(\cdot)$.

\begin{lemma} \label{lem:nodeface}
The representation of Theorem~\ref{thm:basic} can determine in $O(1)$ time
the two nodes connected by an edge, and the two faces separated by an edge.
\end{lemma}
\begin{proof}
The two nodes corresponding to an
edge $i$ in $G$ are simply $node(i)$ and $node(match(S,i))$. The two faces are $face(i)$ and $face(match(S,i))$.

\end{proof}

\subsection{Listing Queries}

Listing the faces bordering a given face (3.d) can be done as the dual of
listing the neighbors of a node (3.c), by exchanging the roles of brackets and
parentheses in Theorem~\ref{thm:basic}. Listing the faces incident on a node 
(3.e) can also be done as a subproduct of Theorem~\ref{thm:basic}. For
each edge $e$ incident on $u$, obtained in counter-clockwise order, we obtain 
the faces $e$ divides using Lemma~\ref{lem:nodeface}. This lists all the faces
incident on $u$, in counter-clockwise order, with the only particularity that
each face is listed twice,
consecutively.
Analogously, given a face identifier $x$, we can list the nodes
found in the frontier of the face (3.f). This query is not exactly the same as
in Theorem~\ref{thm:basic}, because there we must start from an edge bordering
the desired face.

\begin{lemma} \label{lem:listnodesorfaces}
The representation of Theorem~\ref{thm:basic} suffices to list, given a node 
$u$, the faces incident on $u$ in counter-clockwise order from its parent in $T$, each in 
$O(1)$ time, or given a face $x$, the nodes in the frontier of $x$ in
clockwise order
from its parent in $T^*$, each in $O(1)$ time.
\end{lemma}

\subsection{Counting Queries}

Ferres et al.~\cite{FFSGHN18} count the number of edges incident on a 
node $u$ (4.a) in $O(f(m))$ time using $o(m)$ bits, for any $f(m)\in\omega(1)$.
A bitvector of length $O(m)$ with $O(m/f(m))$ 1s marks the nodes with 
degree $f(m)$ or more; this bitvector requires $O(m\log f(m)/f(m))+o(m)$ bits 
in compressed form \cite{RRR07}. For nodes with degree below $f(m)$, they traverse the 
neighbors one by one; for the others, they store the degree explicitly in another bitvector
using $O(m\log f(m)/f(m)) \subseteq o(m)$ further bits.

We can similarly count the number of neighboring nodes or faces, with the
exception that we can reach several times the same node or face as we traverse 
the edges incident on a node. Thus, we need time $O(f(m)\log f(m))$ on nodes 
with degree over $f(m)$ in order to remove repetitions; for higher-degree nodes
we store the correct number explicitly. We then
obtain $O(f(m)\log f(m))$ time using $o(m)$ bits, which
still achieves any time in $\omega(1)$.
By building the structure on the dual of $G$, we count the number of 
edges, nodes, or faces in the frontier of a face $x$ (4.b).

\section{Deciding if Two Nodes/Faces Share an Edge}
\label{sec:neighbors}

\begin{figure}[t]
    \centering
    \begin{subfigure}[b]{0.35\textwidth}
        \centering
        \includegraphics[width=\textwidth]{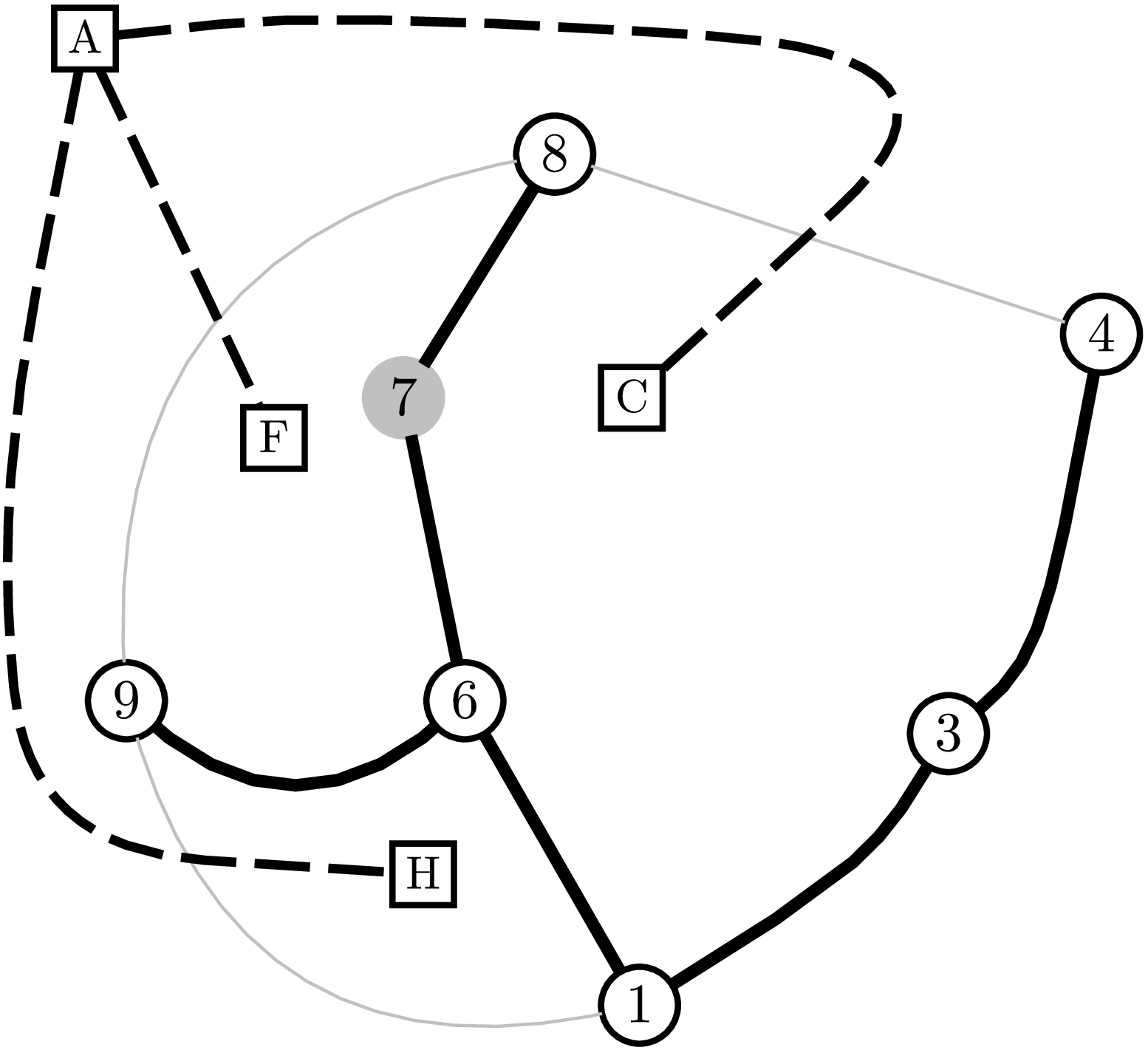}
        \caption{Graph $G'$ with $f(m)=3$}
        \label{fig:gprime}
    \end{subfigure}
    ~ 
    \begin{subfigure}[b]{0.6\textwidth}
        \centering
        \includegraphics[width=\textwidth]{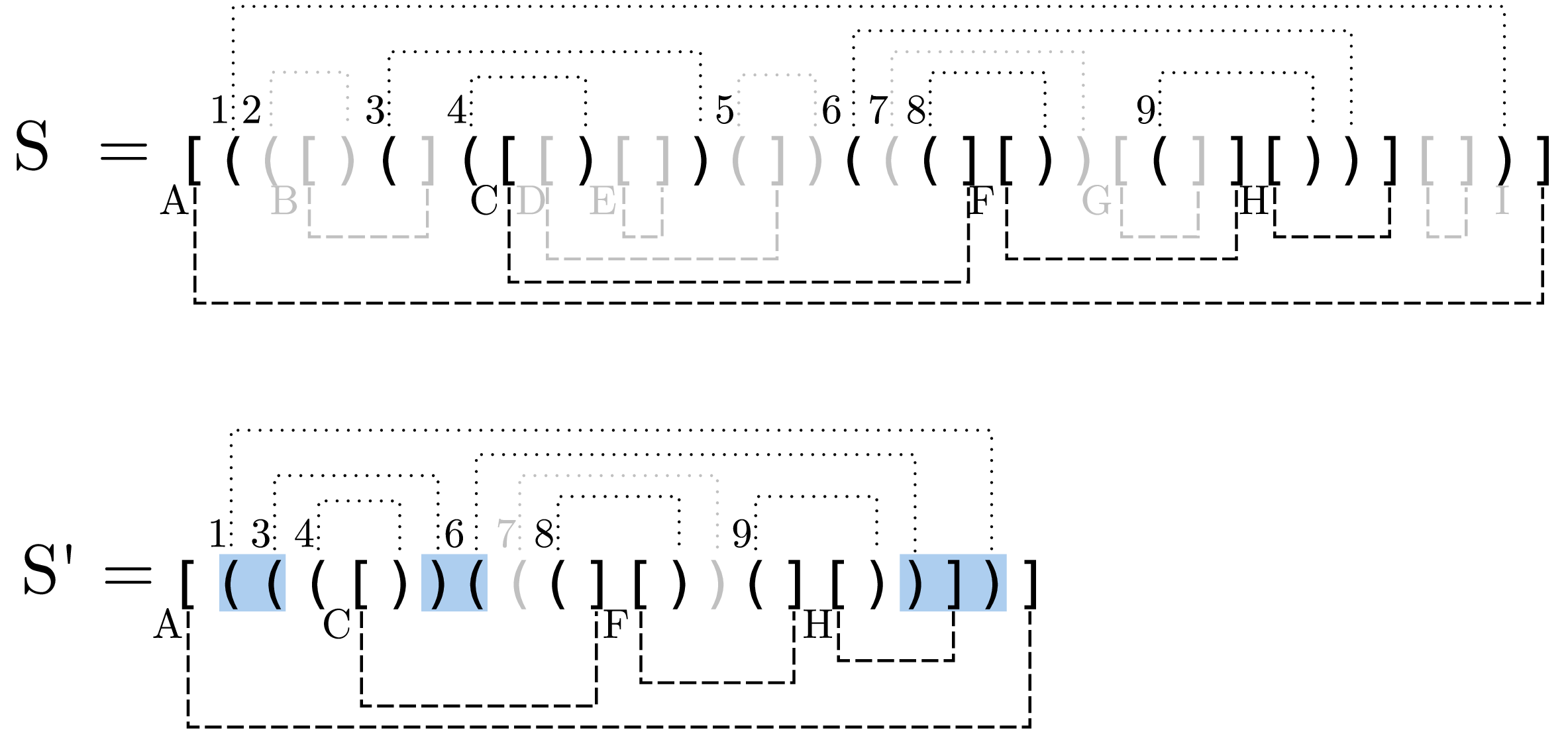}
        \caption{Parentheses/brackets representation of $G'$}
        \label{fig:sprime}
    \end{subfigure}
    \begin{subfigure}[b]{0.35\textwidth}
        \centering
        \includegraphics[width=\textwidth]{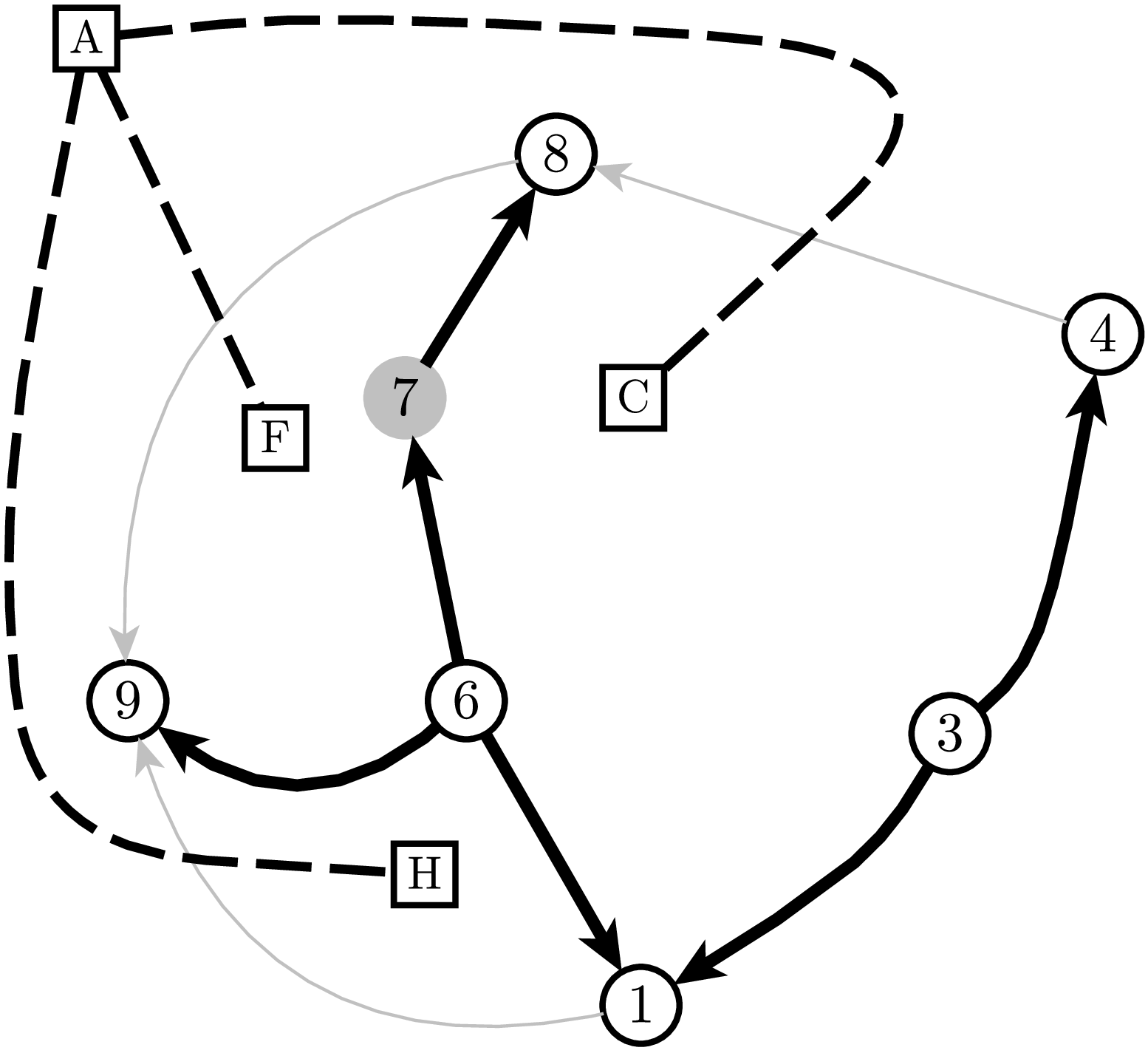}
        \caption{Oriented graph $G'$}
        \label{fig:oriented}
    \end{subfigure}
    ~ 
    \begin{subfigure}[b]{0.6\textwidth}
        \centering
        \includegraphics[width=0.6\textwidth]{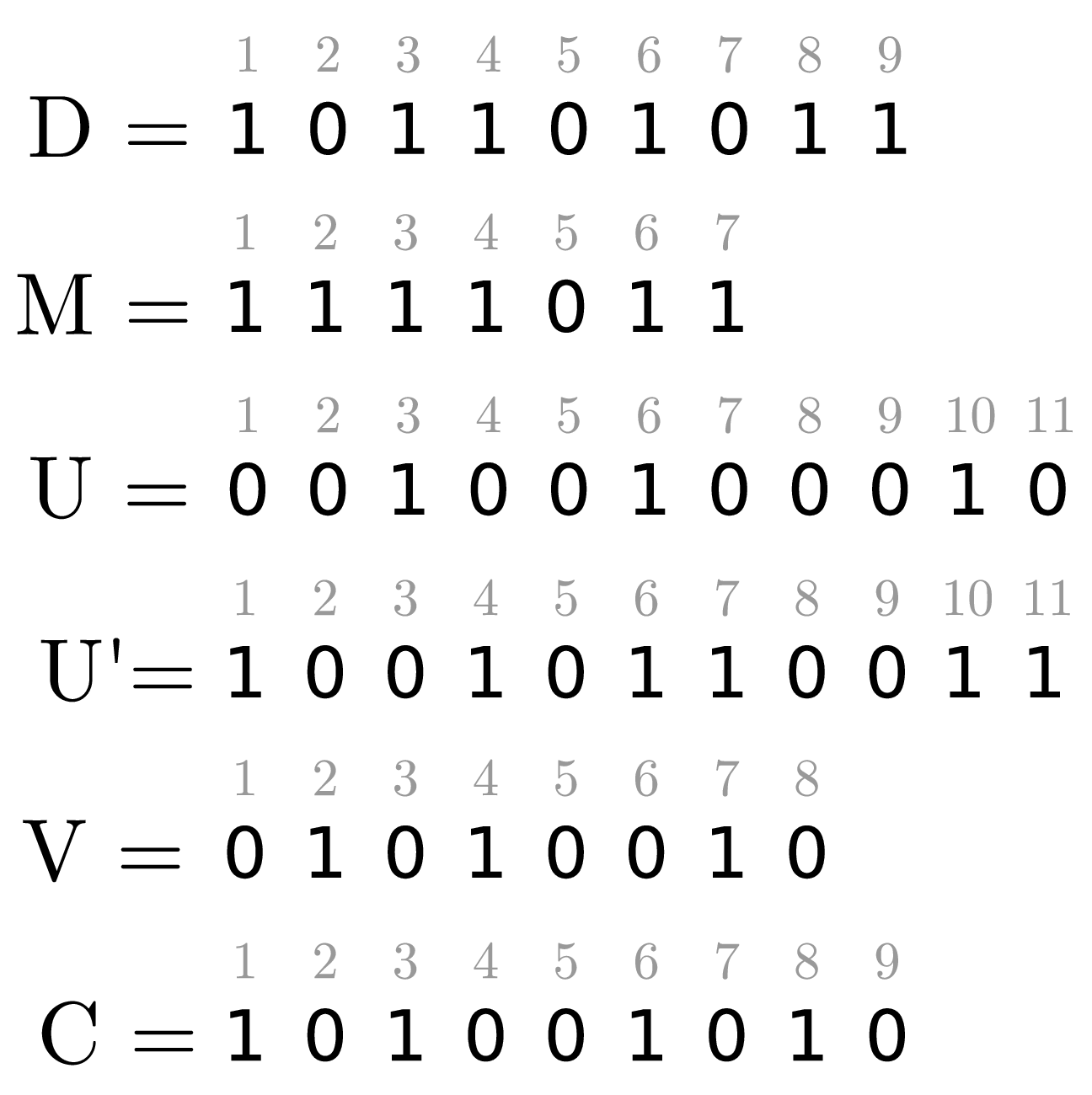}
        \caption{Bitvectors $D$, $M$, $U$, $U'$, $V$ and $C$}
        \label{fig:seqUV}
    \end{subfigure}
    \caption{Graph $G'$ used to support query (1.c). {\bf(\subref{fig:gprime})} Graph $G'$ after applying all the pruning rules, with $f(m)=3$; node $7$ is the only gray node that remains. {\bf(\subref{fig:sprime})} The representation of $G'$. On the top we mark the gray nodes and removed edges in $S$. On the bottom we show the final sequence $S'$; we also highlight in blue the range of parentheses and brackets representing the children of node 1. {\bf(\subref{fig:oriented})} Graph $G'$ with oriented edges. {\bf(\subref{fig:seqUV})} Bitvectors $D$, $M$, $U$, $U'$, $V$, and $C$ used to support query (1.c).}\label{fig:structs_query1c}
\end{figure}

Ferres et al.~\cite{FFSGHN18} show how we can determine if two given nodes
$u$ and $v$ are connected in any time $f(m) \in \omega(\log m)$. First, they
check in constant time if they are connected by an edge of the spanning tree
$T$: one must be the parent of the other. Otherwise, the nodes can be connected
by an edge not in $T$, represented by a pair of brackets. Their idea is to
mark in a bitvector $D[1..n]$ the nodes having $f(m)$ neighbors or more (see Fig.\ref{fig:seqUV} for an example). The
subgraph $G'$ induced by the marked nodes, where they also eliminate self-loops
and multi-edges, has $n' \le 2m/f(m)$ nodes, because at least $f(m)$ edges
are incident on each marked node and each of the $m$ edges are incident on
$2$ nodes. Since $G'$ is planar and simple, it can have only
$m' < 3n' \le 6m/f(m)$ edges.
 They represent $G'$ using adjacency lists, which 
use $o(m)$ bits as long as $f(m) \in \omega(\log m)$. Given two nodes $u$ and
$v$, if either of them is not marked in $D$, they simply enumerate its 
neighbors in time $O(f(m))$ to check for the other node. Otherwise, they map
both to $G'$ using $rank_1(D)$, and binary search the adjacency list of one of
the nodes for the presence of the other, in time $O(\log m) \subseteq o(f(m))$. 
Bitvector $D$ has $n' \le 2m/f(m)$ bits set out of $n \le m+1$ (this second
inequality holds because $G$ is connected), and 
therefore it can be represented using $(2m/f(m))\log(f(m)/2))+O(m/f(m))+o(m)
\subseteq o(m)$ bits while answering $rank$ queries in constant time \cite{RRR07}.\footnote{They do not 
specify how to handle queries of the form $(u,u)$ given that they remove 
self-loops. We can have a bitvector $L[1..n']$ of size $o(m)$ so that, if $D[u]=1$, then there is an edge $(u,u)$ in $G$ iff $L[rank_1(D,u)]=1$.}

We will obtain any time $f(m) \in \omega(1)$ by solving the query on $G'$ in a different way. This
requires a more complex mapping, however, because now we
cannot afford to represent the node identifiers of $G'$ in explicit form within $o(m)$ bits.

In particular, we will not physically remove (all) the unmarked nodes of $G$ to form $G'$; we just {\em paint} the unmarked nodes to signal that they can be removed. We do, instead, remove useless edges not in the spanning tree $T$. More precisely, we start with $G' = G$ and then:
\begin{itemize}
    \item Paint the low-degree nodes $u$ in gray; say the high-degree ones are black.
    \item Remove the edges incident on gray nodes $u$ and not belonging to $T$.
    \item Remove self-loops and multiple edges, though never choosing an edge of $T$.
\end{itemize}

The spanning tree $T'$ of $G'$ is in principle identical to $T$.
Note that the only remaining neighbors of gray nodes are connected by edges in $T'$.
In order to obtain the desired space/time performance the gray nodes must be 
reduced, yet without affecting the traversal order of $T'$ on the remaining nodes. We thus perform the following additional pruning on $G'$ and $T'$:
\begin{enumerate}
\item Consecutive gray siblings in $T'$, if their parent has no other edges between them in $G'$, are merged into one, and their children list are concatenated.
\item A gray node with only one child that is also gray is removed, and its child is connected to its parent.
\item Gray nodes that are leaves in $T'$ are removed.
\end{enumerate}

Fig.~\ref{fig:gprime} shows the graph $G'$ for the graph $G$ of Fig.~\ref{fig:planar}, with $f(m)=3$. 

As seen, $G'$ has $n' \le 2m/f(m)$ black nodes. It has also gray nodes, but by rules (1--3) above, every gray node has a first or a second child in $T'$ that is black (recall that all the edges of a gray node are in $T'$, so rule (1) leaves no consecutive gray children of a gray node in $T'$). Thus, $G'$ has at most $n'$ gray nodes. Further, since $G'$ is simple and contains at most $2n'$ nodes (black or gray), it contains $m' < 6n'$ edges. The length of the sequence $S'$ representing $G'$ is then $2m'+4 < 12n'+4 \le 24m/f(m)+4 \in O(m/f(m)) \subseteq o(m)$.

We use an additional bitvector $M$ that identifies with 1s the black nodes of $T'$, in preorder. Therefore, to map the identifier $u$ of a marked node in $G$ (i.e., $D[u]=1$), we first compute $u'' = rank_1(D,u)$, which is its preorder position among the black nodes of $T'$. We then compute $u' = select_1(M,u'')$ to obtain its node identifier in $T'$. Its opening parenthesis in $S'$ is then at $select_{``(''}(S',u')$. The length of $M$ is at most $2n' \in o(m)$.
An example of the bitvector $M$ can be seen  Fig.~\ref{fig:seqUV}.

The key idea to determine if (black) nodes $u'$ and $v'$ are connected in $G'$ is that one can orient the edges in
a simple planar graph so that every node has outdegree at most $3$ \cite{CE91}. Once $G'$ is oriented, determining whether $(u',v')$ is an edge of $G'$
requires testing whether at most $6$ edges connect $u'$ and $v'$ (i.e., the $3$
edges leaving $u'$ and the $3$ edges leaving $v'$). For instance, Fig.~\ref{fig:oriented} shows a possible orientation for the edges of graph $G'$ of Fig.~\ref{fig:gprime}.

The problem then reduces to finding
each of the out-edges of a node $u'$ fast. We can focus on the edges not in $T'$, 
because we always start checking whether $(u,v)$ or $(v,u)$ are in $T$, and the edges between black nodes of $T'$ also belong to $T$. 
We will find in constant time the (at most) $3$ brackets that represent out-edges
in the top level of the sequence $(\cdots)$ that describes $u'$ (we say that those brackets are {\em marked}). Recall that 
this sequence contains the sub-sequences $(\cdots)$ of the children $u_1,\ldots,
u_k$ of $u'$ in $T'$, interspersed with brackets. 

We proceed as follows. We define a bitvector $U$ where we traverse $T'$
in preorder and, for each black node $u'$ with $k$ children $u_1,
\ldots,u_k$, we add $k+1$ bits. The first bit is $1$ iff there are marked
brackets between the opening parenthesis of $u'$ and the opening parenthesis of $u_1$ (or, if $u'$
has no children, between the opening and closing parentheses of $u'$). For 
$1 < j \le k$, the $j$th of the $k+1$ bits is $1$ iff there are marked brackets
between the closing parenthesis of $u_{j-1}$ and the opening parenthesis of 
$u_j$.  Finally, the $(k+1)$th bit is a $1$ iff there are marked brackets
between the closing parenthesis of $u_k$ and that of $u'$. Since $T'$ has at most $2n'$ nodes and at most $2n'-1$ of those are children of some node, the length of $U$ is 
$< 4n' \in O(m')$. A second bitvector, $U'$, marks with $1$s the
first of the $k+1$ bits of each node described in $U$. Finally, we use a third bitvector
$V$ of length $O(m')$, where $V[i]=1$ means that the $i$th bracket, left to 
right in $S'$, is marked. For example, the bitvectors $U$, $U'$ and $V$ for the graph of Fig.~\ref{fig:oriented} are shown in Fig.~\ref{fig:seqUV}.

To find the out-neighbors of a node $u'$, we first find its area $U[p..p']$,
with $p=select_1(U',u')$ and $p' = select_1(U',u'+1)-1$. We now find the (up to 
$3$) positions $p_i \in [p..p']$ where $U[p_i]=1$, with 
$p_i = select_1(U,rank_1(U,p-1)+i)$, stopping when $p_i > p'$.
For each of those $p_i$, we must search the area of brackets that
lie between the $(p_i-p)$th and the $(p_i-p+1)$th children of $u'$. Let $u'$
have $k$ children. The area between the $0$th and the $1$st children refers
to $S'[select_{``(''}(S',u')+1..select_{``(''}(S',u'+1)-1]$. The area between
the $i$th and the $(i+1)$th children, for $1 \le i \le k$, refers to
$S'[match(S',child(S',u',i))+1..child(S',u',i+1)]$, where we extend the operation
$child$ to operate on the parentheses of the sequence $S'$ as follows:
\[ child(S',u',i) = select_1(A',child(B',select_0(B',u'),i)),
\]
where we assume $S'$ is represented with bitvectors $A'$, $B'$, and $(B^*)'$.
Finally, the area between the $k$th and the $(k+1)$th children of $u'$ refers to 
$S'[match(S',child(S',u',k))+1..match(S',u')-1]$.

Let $S'[q..q']$ be any such area of $S'$, which is composed of only brackets. We map $q$ and $q'$ to $V$ with $r = rank_0(A',q)$ and $r' = r+(q'-q)$, so we must enumerate the (up to 3) 1s in $V[r..r']$. Those are $r_i = select_1(V,rank_1(V,r-1)+i)$, stopping when $r_i > r'$. The corresponding marked brackets are $q_i = r_i + (q-r)$. Each position $S'[q_i]$ corresponds to an edge that must be 
tested to see if it is incident on $v'$, in constant time with query (2.a).
We then analogously check if the up to $3$ out-neighbors of $v'$ are incident on $u'$.

If we further wish to retrieve the positions $S[b..b']$ of a pair of brackets 
that connect $u'$ and $v'$, when the edge does not trivially belong to 
$T$, we enrich our
structure with bitvector $C$, which
tells which face identifiers of $G$ (i.e., ranks of opening brackets) survive in 
$G'$. See the bottom of Fig.~\ref{fig:seqUV} for an example.
Once we find that $u'$ and $v'$ are neighbors 
connected by the edge $S'[x..x']=[\cdots]$, we have that the opening bracket 
number $b' = rank_{``[''}(S',x)$ connects 
them in $G'$. We then identify the edge 
in $G$ with $b=select_1(C,b')$, and $mate(b)$, in $O(1)$ additional time.
The length of bitvector $C$ is less than $m$ and it has less than $m'$ 1s,
thus it can be represented in $O(m\log f(m)/f(m))+o(m) \subseteq o(m)$ bits 
\cite{RRR07}. 

We thus solve query (1.c) with $o(m)$ extra bits of space and $O(f(m))$ time, for any $f(m) \in \omega(1)$.

\begin{lemma} \label{lem:neigh}
The representation of Theorem~\ref{thm:basic} can be enriched with $o(m)$ bits
so that we can determine whether two nodes are connected
in any time in $\omega(1)$.
\end{lemma}

\subsection{Determining Adjacency of Faces}

By exchanging the interpretation of parentheses and brackets, the same sequence 
$S$ represents the dual $G^*$ of $G$, where the roles of nodes and faces are 
exchanged. We can then use the same solution of Lemma~\ref{lem:neigh} to 
determine whether two faces are adjacent (1.d).
 We do not explicitly store the sequence $S^*$ representing $G^*$,
since we can simulate it using $S$. We do, instead, build a structure on $S^*$ 
analogous to the one we built on $S$, creating sequence $(S^*)'$ and its auxiliary bitvectors.
 This time, the input to the query are the ranks of the 
opening brackets representing both faces (i.e., node identifiers in $G^*$).
We then solve query (1.d).

\begin{lemma} \label{lem:dualneigh}
The representation of Theorem~\ref{thm:basic} can be enriched with $o(m)$ bits
so that we can determine whether two faces are adjacent
in any time in $\omega(1)$.
\end{lemma}

%

\section{Determining Incidence of a Face in a Node}
\label{sec:nodesfaces}

Given a node $u$ and a face $x$, the problem is to determine whether $x$ is 
incident on $u$ (2.c). Since with Lemma~\ref{lem:listnodesorfaces} we can list 
each face incident on $u$ in constant time, or each node bordering $x$ in 
constant time, we can use a scheme combining those of Lemmas~\ref{lem:neigh} 
and \ref{lem:dualneigh}: If $u$ has less than $f(m)$ neighbors, we traverse 
them looking for $x$. Otherwise, if $x$ has less than $f(m)$ bordering nodes, 
we traverse them looking for $u$. 
We now show how to handle the remaining case.

We define a graph $G^\#$ where we add additional nodes representing {\em selected} faces of $G$, that is, those having at least $f(m)$ nodes in their frontier.
The queries that are not handled by enumeration in $G$ will become a node neighbor query on $G^\#$, and solved as in Lemma~\ref{lem:neigh}. The graph $G^\#$, which will have $O(m)$ edges, will not be represented directly. 

Concretely, $G^\#$ adds to $G$ a new node $v(x)$ per selected face $x$, as well as new edges connecting $v(x)$ with all nodes in the frontier of $x$. Note $G^\#$ is planar because we can draw $v(x)$ inside the face $x$. There are at most $2m/f(m)$ selected faces $x$, because each has at least $f(m)$ edges in its frontier and each edge is in the frontier of two faces.
The graph $G^\#$ contains $n' \le n + 2m/f(m) = O(m)$ nodes and $m' \le 3m$ edges (each selected face limited by $j$ edges of $G$ adds $j$ new edges in $G^\#$, and each edge of $G$ limits two faces). 

\begin{figure}[t]
    \centering
    \begin{subfigure}[b]{\textwidth}
        \centering
        \includegraphics[width=0.5\textwidth]{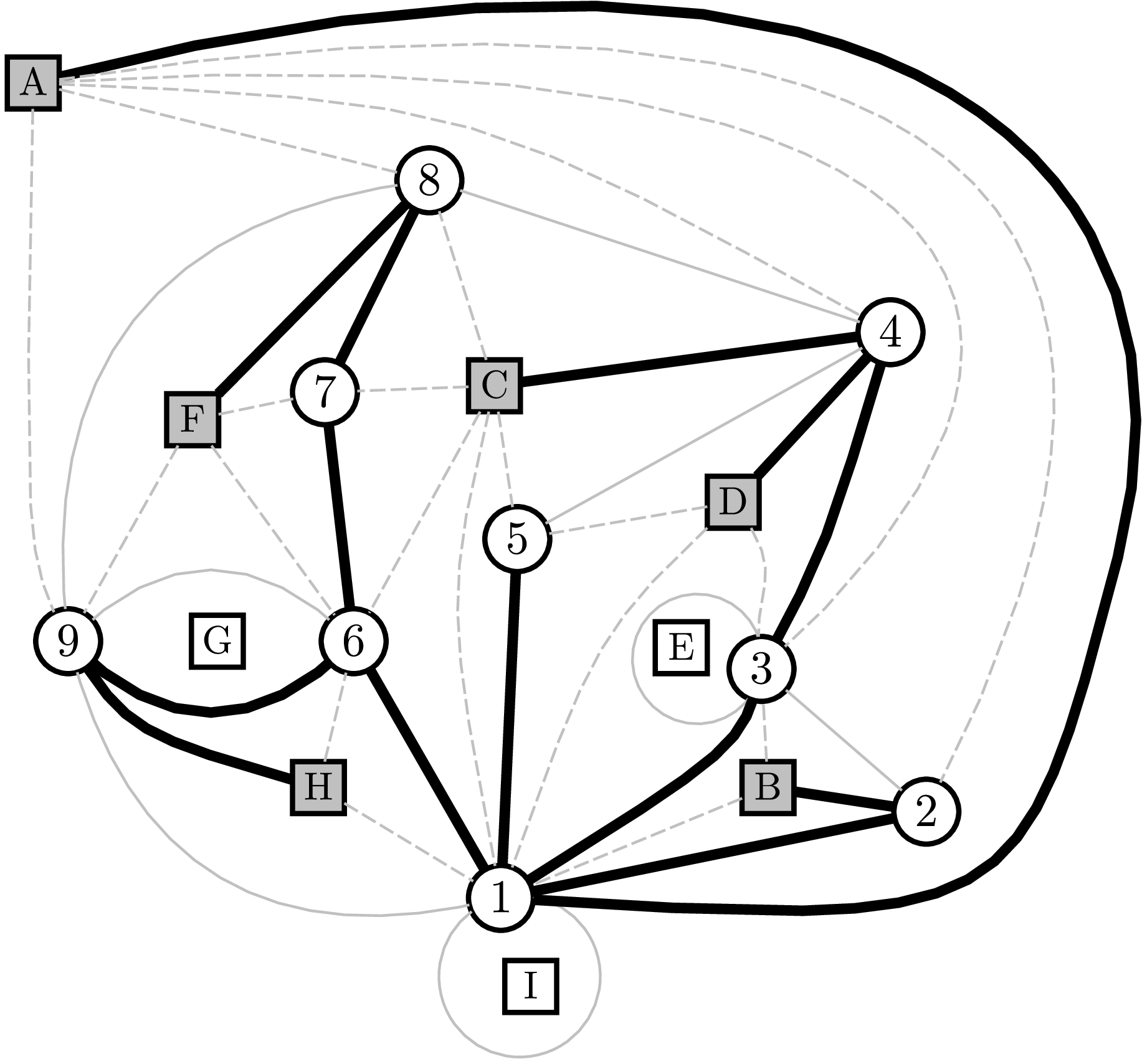}
        \caption{Graph $G^\#$ with $f(m)=3$}
        \label{fig:gnumber}
    \end{subfigure}
    ~ 
    \begin{subfigure}[b]{\textwidth}
        \centering
        \includegraphics[width=\textwidth]{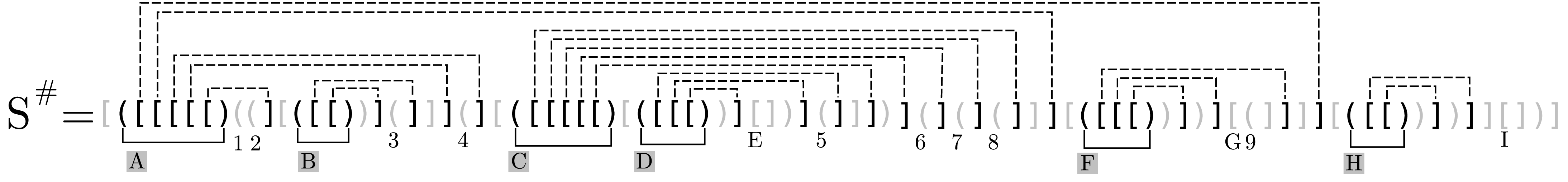}
        \caption{Sequence $S^\#$ generated from the graph of Fig.~\ref{fig:gnumber}}
        \label{fig:snumber}
    \end{subfigure}
    \caption{Graph $G^\#$ used to support query (2.c). {\bf (\subref{fig:gnumber})} Graph $G^\#$ with $f(m)=3$. The selected faces are painted in gray and the spanning tree $T^\#$ is represented with thick edges. {\bf (\subref{fig:snumber})} Sequence $S^\#$ obtained from the traversal of $T^\#$. The parentheses and brackets added to those in $S$ are in black, so $S$ is the gray subsequence of $S^\#$.}\label{fig:structs_query2c}
\end{figure}

A spanning tree $T^\#$ for 
$G^\#$ is built by extending the spanning tree $T$ of $G$ with leaves $v(x)$, as
follows. Consider the traversal of $G$ that defines the spanning trees
$T$ and $T^*$. Let $(u,v) \not\in T$ be the edge in the frontier of a selected face $x$ 
where the traversal of $T^*$ first reaches the face $x$, that is, when edge 
$(y,x)$ is added to $T^*$ for some face $y$. Right after visiting the edge $(u,v)$, we add a new leaf node 
$v(x)$, as a child of $u$, to $T^\#$. We also add the edges $(x,w)$ for the other 
nodes $w$ in the frontier of face $x$; those edges will not belong to $T^\#$.
An example of $G^\#$ and $T^\#$ is shown in Fig.~\ref{fig:gnumber}.

To generate the sequence $S^\#$ representing $G^\#$ we traverse $T^\#$
starting from the edge that connects the outer face with the starting node that generates the sequence $S$. After that, the traversal follows the
same order of $T$. When we reach an original selected face $x$ (which in $G^\#$ is 
partitioned into $j$ triangles), we will visit the edge $(u,v(x)) \in T^\#$ right 
after $(u,v) \not\in T^\#$. We will then traverse the other $j-1$ edges incident
on $v(x)$, none of which is in $T^\#$, and all of which are visited for the first
time because we had not entered face $x$ before. Therefore, the $``[''$ that
represents $(u,v)$ in $S^\#$ will be immediately 
followed by $(~[^{j-1} )$, and then the normal layout of $T$ will follow. 
Those opening brackets will be closed later along the traversal. Fig.~\ref{fig:snumber} shows the sequence $S^\#$ obtained after traversing the spanning tree $T^\#$ of Fig.~\ref{fig:gnumber}, starting from the edge $(1,A)$. For instance, face $C$ is bounded by the nodes 1, 5, 4, 8, 7 and 6, which are represented by $(~[~[~[~[~[~)$ in Fig.~\ref{fig:snumber}.

This implies that every opening bracket in $S$ representing a selected face $x$, 
is immediately followed in $S^\#$ by an opening parenthesis
corresponding to the node $v(x)$ we created for the face. That is, selected faces $x$ in $S$ are in the same order of added nodes $v(x)$ in $S^\#$. Further, the nodes of $G$ are in the same order in $S$ and $S^\#$. We exploit this correspondence to map nodes and faces from $G$ to nodes of $G^\#$ by using the following bitvectors:
\begin{itemize}
    \item The same bitvector $D[1..n]$ of Section~\ref{sec:neighbors}, where $D[u]=1$ iff node $u$ of $G$ has at least $f(m)$ neighbors.
    \item A bitvector $E[1..m-n+1]$, where $E[x]=1$ iff face $x$ of $G$ has at least $f(m)$ nodes (or edges) in its frontier.
    \item A bitvector $R$ where $R[x']=1$ iff node $x'$ of $G^\#$ is one of the nodes $v(x)$ we added to $G$ in order to form $G^\#$.
\end{itemize}

If $D[u]=1$, we map it to node $u' = select_0(R,u)$ in $G^\#$ (note that all the nodes in $G$ appear in $G^\#$). If $E[x]=1$, we map it to node $x' = select_1(R,rank_1(E,x))$ (only the selected faces in $G$ appear as new nodes in $G^\#$). Bitvectors $D$, $E$, and $R$ are of length $O(m)$ and have $O(m/f(m))$ 1s, so they can be represented within $O(m\log f(m)/f(m)) + o(m) \subseteq o(m)$ bits \cite{RRR07}. 

If $u$ has at least $f(m)$ neighbors and $x$ is limited by at least $f(m)$ nodes, we map node $u$ and face $x$ to nodes $u'$ and $x'$ in $G^\#$ as explained, and determine if they are neighbors. Note that, since $x$ has at least $f(m)$ nodes in its frontier, node $x'$ has at least $f(m)$ neighbors. Node $u'$ also has at least $f(m)$ neighbors in $G^\#$, because it had in $G$ and it can only get further neighbors in $G^\#$. We then build on $S^\#$ the structures of Lemma~\ref{lem:neigh}, without explicitly representing $S^\#$, so that we can map $u'$ and $x'$ to the reduced sequence $(S^\#)'$ and solve the query in there. This adds $o(m)$ bits of space and completes the query in any time in $\omega(1)$.

\begin{lemma} \label{lem:nodesfaces}
The representation of Theorem~\ref{thm:basic} can be enriched with $o(m)$ bits
so that, given a node $u$ and a face $x$, it answers whether
$u$ is in the frontier of $x$ in any time in $\omega(1)$.
\end{lemma}

\section{Determining Indirect Connections}
\label{sec:inters}

To handle queries (5.a) and (5.b), we reuse the idea of selecting a
subgraph where the query cannot be solved in time $O(f(m))$ and storing a suitable speed-up structure for those cases. This time, however, the idea leads to a much 
higher time complexity. We later obtain constant time by relaxing the space requirement to $O(m)$ bits.

Let us first consider determining if two nodes are in the border of the same (unknown) face.
Given two nodes $u$ and $v$, if either has less than $f(m)$ neighbors we can
traverse its incident faces one by one and, for each face $x$, use
Lemma~\ref{lem:nodesfaces} to determine if $x$ is incident on the other node
in time $\omega(1)$. For all the pairs of nodes $(u,v)$ where both have
$f(m)$ neighbors or more, we store a binary matrix telling whether or not they
lie on the same face. This requires $(2m/f(m))^2$ bits, which is $o(m)$ for any
$f(m) \in \omega(\sqrt{m})$. Thus we can solve query (5.a) and, by duality,
query (5.b), in any time in $\omega(\sqrt{m})$.

\begin{lemma} \label{lem:sqrt}
The representation of Theorem~\ref{thm:basic} can be enriched with $o(m)$ bits
so that, given two nodes or two faces, it answers in $O(f(m))$ time whether
they share a face or a node, respectively, for any $f(m) \in 
\omega(\sqrt{m})$.
\end{lemma}

If we want to know the identity of the shared face (or, respectively, node),
this can be stored in the matrix, which now requires $O((m/f(m))^2\log m)$
bits. We can then reach any time in $\omega(\sqrt{m\log m})$.

\subsection{Constant Times with $O(m)$ Bits of Space}

As explained, those operations requiring $\omega(1)$ time in Table~\ref{tab:model} automatically become $O(1)$ if we use $f(m) \in O(1)$. In exchange, the space becomes $4+\epsilon$ bits of space for any desired constant $\epsilon>0$. We now show that queries (5.a) and (5.b) can also be solved in $O(1)$ time if we raise the space to $O(m)$ bits.\footnote{In the conference version \cite{FNS19}, we incorrectly conjectured
that this problem was intersection-hard \cite{CP10,PR14} even using $O(m\log m)$ bits of space.}
Let us focus on the query of Lemma~\ref{lem:sqrt} (5.a); we then obtain (5.b) by duality. Our solution is inspired in a 
(non-compact) data structure for constant-time bounded shortest distance 
queries on planar graphs \cite{KK06}.

\begin{figure}[t]
    \centering
    \begin{subfigure}[b]{0.45\textwidth}
        \centering
        \includegraphics[width=\textwidth]{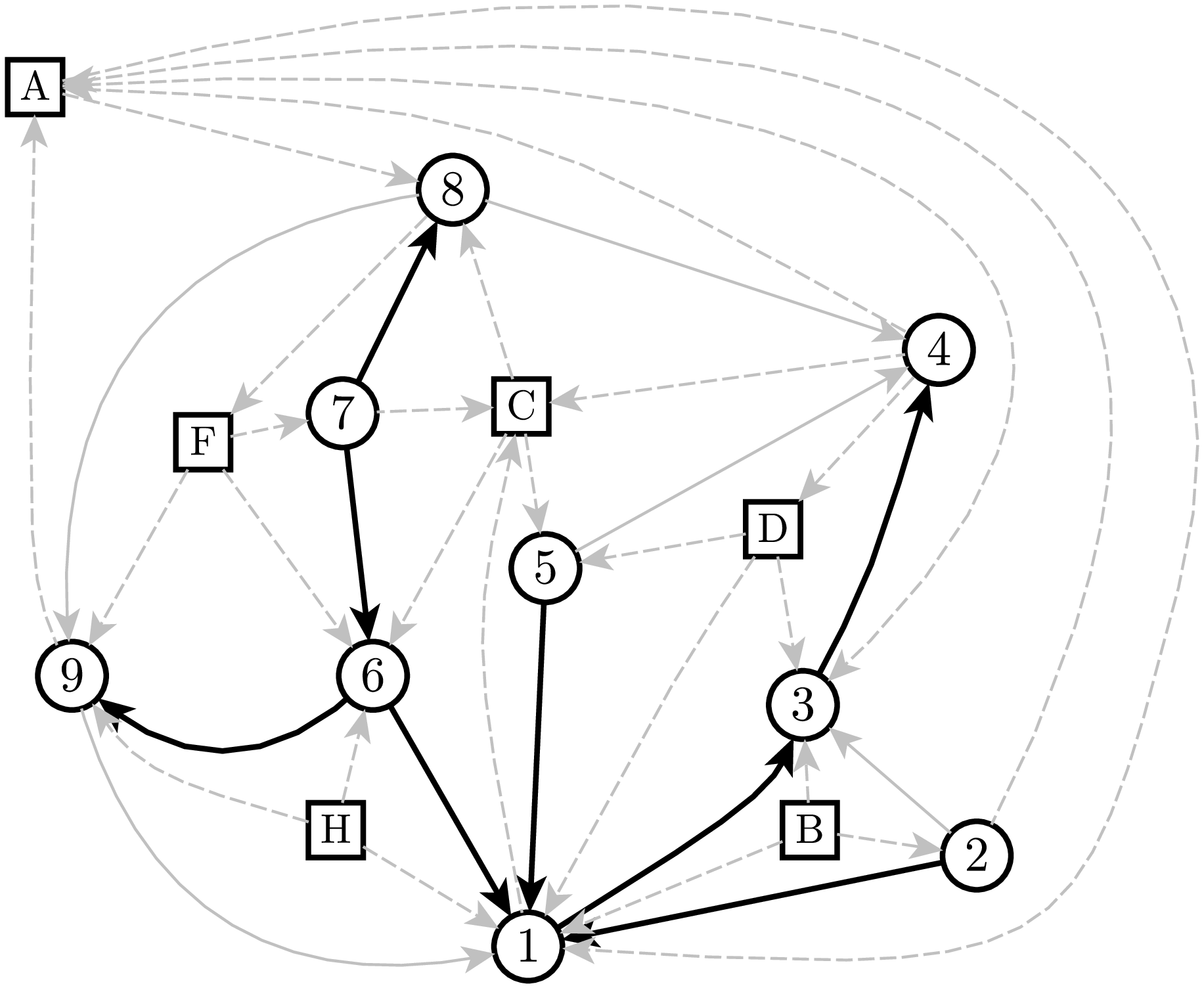}
        \caption{Graph $G^\#$ with oriented edges}
        \label{fig:gnumber_oriented}
    \end{subfigure}
    ~ 
    \begin{subfigure}[b]{0.45\textwidth}
        \centering
        \includegraphics[width=\textwidth]{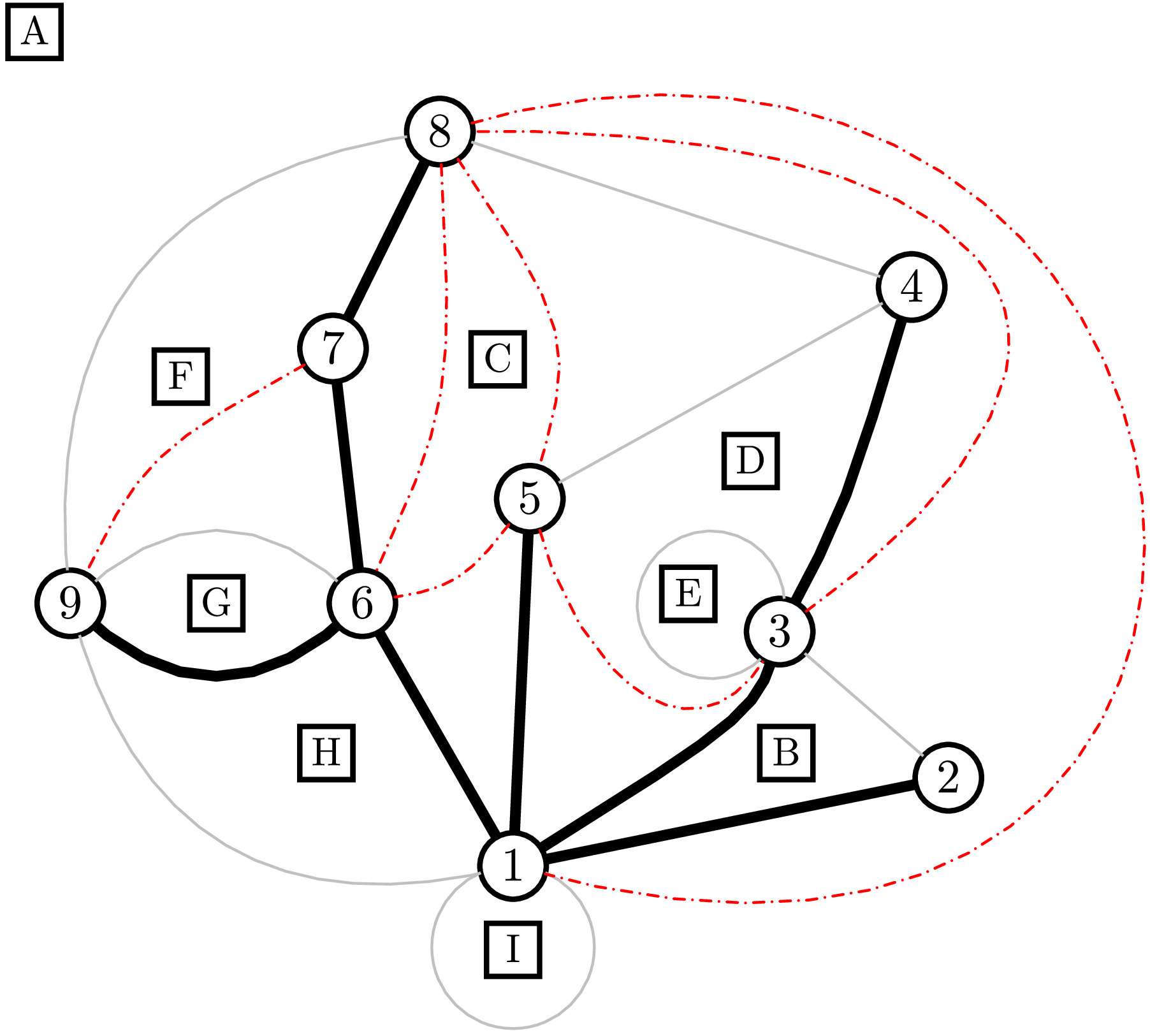}
        \caption{Graph $G^{\$}$}
        \label{fig:gdollar}
    \end{subfigure}
    \caption{Graphs $G^\#$ and $G^\$$ used to support query (5.a). {\bf (\subref{fig:gnumber_oriented})} Graph $G^\#$. Solid edges represent edges of the original graph $G$ (Fig.~\ref{fig:planar}), and dashed ones represent edges connecting faces with their bordering nodes. The spanning tree $T$ of the original graph is in black. {\bf (\subref{fig:gdollar})} Graph $G^\$$. Red edges represent the new edges connecting nodes linked by two out-edges of the node induced by their common face in $G^\#$.}
    \label{fig:structs_query5a}
\end{figure}

Consider the graph $G^\#$ of Section~\ref{sec:nodesfaces}, with the following changes:
\begin{itemize}
    \item We start with a copy of $G$ and then remove self-loops and multiple edges (not those that belong to $T$). Removing self-loops and multiple edges is irrelevant for the query (5.a). 
    \item We select {\em all} the remaining faces of $G$.
    \item We represent $G^\#$ explicitly (so we use $O(m)$ bits of space).
    \item We orient the edges of $G^\#$ as in Section~\ref{sec:neighbors}. Since $G^\#$ is simple, the out-degree of every node can be made at most $3$. 
\end{itemize}

If nodes $u$ 
and $v$ are both in the frontier of some face mapped to node $x$ in $G^\#$, then the following
configurations of the edge orientations are possible in $G^\#$:
$u \rightarrow x \rightarrow v$, $u \rightarrow x \leftarrow v$, 
$u \leftarrow x \leftarrow v$, and $u \leftarrow x \rightarrow v$. 
The first three are easily verified in constant time using the structures of
Section~\ref{sec:nodesfaces}. For example, for 
$u \rightarrow x \rightarrow v$ we
check the (up to) $3$ out-neighbors $x_i$ of $u$ and, for those $x_i$ that are
faces in $G$ (which is verified in bitvector $R$), we check their (up to)
$3$ out-neighbors $v_{i,j}$, to see if $v_{i,j}=v$.

The difficult case is $u \leftarrow x \rightarrow v$, because we should 
start our quest from the unknown node $x$. Fortunately, inside the face $x$ represents
in $G$, there can be at most $3$ nodes that are out-neighbors of $x$ in
$G^\#$. We then create another extended version of the original graph $G$,
which we call $G^\$$, where we additionally connect those $3$ nodes inside each face $x$,
thereby drawing a triangle inside the face (thus $G^\$$ is planar). Thus, there is a configuration
$u \leftarrow x \rightarrow v$ iff $(u,v)$ is one of those edges of the 
triangles we have added. 
Fig.~\ref{fig:gdollar} shows an example for the graph $G^\$$, based on the oriented edges of Fig.~\ref{fig:gnumber_oriented}. For instance, the cases 
$8 \leftarrow A \rightarrow 1$ and $5 \leftarrow C \rightarrow 6$ cause the insertion of the edges $(1,8)$ and $(5,6)$, as shown in Fig.~\ref{fig:gdollar}.
It is clear that $G^\$$ has $O(m)$ edges and that
we can define a spanning tree $T^\$$ for it that is identical to that of $T$,
letting all the added edges be not in $T^\$$. We then represent $G^\$$ as in
Lemma~\ref{lem:neigh} (with $f(m) \in O(1)$) and verify the configuration 
$u \leftarrow x \rightarrow v$ by asking if $u$ and $v$ share an edge in
$G^\$$. Note that this query can return an edge that belongs to $G$, but
in this case it is also true that $u$ and $v$ border the same face.
By also considering duality, we have the following result.

\begin{lemma} \label{lem:inters-constant}
The representation of Theorem~\ref{thm:basic} can be enriched with $O(m)$ bits
so that, given two nodes or two faces, it answers in $O(1)$ time whether
they share a face or a node, respectively.
\end{lemma}

\section{Conclusions}

We built on a recent extension \cite{FFSGHN18} of Tur\'an's representation
\cite{Turan1984} for plane embeddings so as to support a rich set of topological queries within succinct space, $4m+o(m)$ bits for an $m$-edge
embedding. Though it exceeds the asymptotically optimal space of $3.58m+o(m)$
bits, this representation is particularly attractive to handle the topological
model because it regards the graph and its dual symmetrically, thereby enabling
a number of queries relating nodes, edges, and faces. 

Starting with an improved solution to determine if two nodes are neighbors, 
we exploit analogies and duality to support most of the operations in any time in $\omega(1)$. We then relax our space requirements 
to $O(m)$ bits, showing that in this case we can represent variants of the
graph that allow us support all the desired queries (on the original graph)
in $O(1)$ time.

An interesting challenge is whether we can support bounded distance
queries (bounded meaning that only distances up to some constant $k$ are
distinguished) efficiently and within $O(m)$ bits of space. For $k=2$, a relatively obvious variant of Lemma~\ref{lem:sqrt} yields any time in $\omega(\sqrt{m})$ within $4m+o(m)$ bits of space. We cannot use an analogous to the $O(m)$-bits construction of Lemma~\ref{lem:inters-constant} to obtain constant time, however, because the resulting graph $G^\$$ could be non-planar.
Kowalik and Kurowski \cite{KK06} show that this query can be solved in
constant time using $O(m\log m)$ bits, that is, with a classical non-compact
representation. They use the same idea of orienting the edges and are left
with the hard subproblem of the configuration $u \leftarrow x \rightarrow v$, for which we built $G^\$$.
They handle this case by adding the edges $(u,v)$ explicitly, which in general
make the graph non-planar. They show, however, that the resulting graph is the
union of a constant number of planar graphs, which can then be queried one by one. Our problem to
obtain $O(m)$ bits from this idea is how to track the node identifiers across
those planar graphs, which can have very different spanning trees.

%
%

\bibliography{paper}

\end{document}